\documentclass[lnbip]{svmultln}

\usepackage{todonotes}
\usepackage{tabularx}
\newcolumntype{L}[1]{>{\raggedright\arraybackslash}p{#1}} 
\newcolumntype{C}[1]{>{\centering\arraybackslash}p{#1}} 
\newcolumntype{R}[1]{>{\raggedleft\arraybackslash}p{#1}} 
\usepackage{makeidx}  
\usepackage{footnote}
\usepackage{url}
\usepackage[para,online,flushleft]{threeparttable}
\begin{document}
\mainmatter              
\title{What Do Practitioners Vary in Using Scrum?}
\titlerunning{Variations of Scrum in Practice}  
%
\author{Philipp Diebold\inst{1} \and Jan-Peter Ostberg\inst{2} \and Stefan Wagner\inst{2} \and Ulrich Zendler\inst{2}}
\authorrunning{Philipp Diebold et al.}   
%
\tocauthor{Ivar Ekeland, Roger Temam, Jeffrey Dean, David Grove,
Craig Chambers, Kim B. Bruce, Elisa Bertino}
\institute{Fraunhofer Institute for Experimental Software Engineering IESE, Germany\\
\email{philipp.diebold@iese.fraunhofer.de}
\and
University of Stuttgart, Germany\\
\email{jan-peter.ostberg|stefan.wagner|ulrich.zendler@iste.uni-stuttgart.de}}

\maketitle              

\begin{abstract}        
\textit{Background}: Agile software development has become a popular way of developing software. Scrum is the most frequently used agile framework, but it is often reported to be adapted in practice. 
\textit{Objective}: Thus, we aim to understand how Scrum is adapted in different contexts and what are the reasons for these changes. 
\textit{Method}: Using a structured interview guideline, we interviewed ten German companies about their concrete usage of Scrum and analysed the results qualitatively.  
\textit{Results}: All companies vary Scrum in some way. The least variations are in the Sprint length, events, team size and requirements engineering. Many users varied the roles, effort estimations and quality assurance.
\textit{Conclusions}: Many variations constitute a substantial deviation from Scrum as initially proposed. For some of these variations, there are good reasons. Sometimes, however, the variations are a result of a previous non-agile, hierarchical organisation.

\vspace{.4cm}
\tiny
The final publication is available at Springer via \url{http://dx.doi.org/10.1007/978-3-319-18612-2\_4}.
\end{abstract}
\section{Introduction}
Nowadays, agile software development has become a common way of developing software, especially in the information systems domain. A survey on agile development~\cite{versionOne} shows that, although there are many agile process frameworks, only few are regularly used: \emph{Scrum}, \emph{Extreme Programming} (XP) and \emph{Kanban}. Scrum is the most frequently used agile process framework with more than 70\% of the answering companies using it~\cite{versionOne}.
Yet, only 55\% use ``pure'' Scrum as it has been initially described. Practitioners apply a combination of Scrum and other approaches or processes, e.g.\ XP or Kanban, as well as adaptations. Ken Schwaber, one of the Scrum inventors, states that around "75\% of companies that claim using Scrum, do not really use Scrum"~\cite{SchwaberGiants}.

Therefore, our research objective is to understand these variations in the application of Scrum in practice. We investigate \textbf{which} variations were introduced and \textbf{why} they are used.
To do so, we interviewed employees of ten German software companies from different domains and with different sizes and analysed the answers qualitatively.


\section{Scrum Background}
\label{sec:background}

To be able to identify \emph{variations}, we need to establish what the \emph{standard} is. We use the ``Scrum Guide'' \cite{sutherland2011Scrum} as our basis for comparison.
Thus, we will summarise the  aspects that are most important for this paper: 

The roles involved in Scrum are: \textbf{Scrum Master}, \textbf{Product Owner} and \textbf{Development Team}.
The Scrum Master is responsible for the team sticking to the rules of Scrum and for organising the events. It is his or her task to introduce changes to optimise the productivity of the Development Team.
The Product Owner is the interface between the Development Team and the stakeholders of the project. It is his or her task to collect all requirements and add them to the \textbf{Product Backlog}, the list of known requirements and related tasks. The Product Owner has to prioritise the requirements in the Product Backlog. She or he is the only one authorised to change the Product Backlog, and ``the Product Owner is one person, not a committee''~\cite{sutherland2011Scrum}. The Development Team has a size of three to nine developers who are self-organising and cross-functional.

The product is developed in iterations called \textbf{Sprints} taking two to four weeks with a fixed length (that could vary over the teams). A Sprint can only be abandoned by the Product Owner if the aim of the Sprint does not match the aim of the project anymore. At the end of each Sprint, a releasable working (software) product is available.

Each Sprint contains the following events:
\begin{itemize}
\item The \textbf{Sprint Planning} defines the aim of the Sprint: The Product Owner presents the backlog items with the highest priority, and the Team estimates how many of them can be accomplished in the next Sprint. This results in the \textbf{Sprint Backlog} containing all requirements the team committed to accomplish.
\item During the Sprint, the Development Team holds a \textbf{Daily Scrum} of 15 minutes maximum supervised by the Scrum Master.
In this event three questions are answered: \textit{What have I accomplished yesterday to fulfil the Sprint aim? What will I do today to approach the Sprint aim? Did I encounter a problem which could interfere with the progress? }
\item In the \textbf{Sprint Review}, at the end of each Sprint, the Sprint results are presented to the stakeholders and accepted based on a common definition of ``Done''.
The stakeholders give feedback about the new increment and further progress is discussed.
\item In the \textbf{Sprint Retrospective}, the Development Team reflects about the Sprint to detect problems and develop solutions.
\end{itemize}

\section {Case Study Design}

\subsection{Research Questions}

Our research objective is to better understand the variations of Scrum in practice and the reasons for these variations. Thus, our study goal is:

\begin{center}
\fbox{\parbox{12cm}{\textit{Analyse the \textbf{Scrum framework} to explore its \textbf{industrial usage} with respect to its \textbf{variations} from the perspective of \textbf{practitioners}.}}}
\end{center}

We broke down this research goal, which still covers a wide area, into research questions (RQ) for a detailed analysis. Based on the description of Scrum (Section~\ref{sec:background}) as the standard for comparison, we ended up with the following research questions focusing on the variations and reasons of their application to Scrum:

\begin{itemize}
\setlength{\itemsep}{2pt}
	\item \textbf{RQ1:} What and why do they vary in the \textbf{Development Team}?
	\begin{itemize}
	\setlength{\itemsep}{1pt}
		\item \textbf{RQ1.1:} What and why do they vary in the \textbf{role} of \textbf{Product Owner}? 
		\item \textbf{RQ1.2:} What and why do they vary in the \textbf{role} of \textbf{Scrum Master}? 
	\end{itemize}
	\item \textbf{RQ2:} What and why do they vary in the \textbf{Sprints}? 
	\item \textbf{RQ3:} What and why do they vary in the \textbf{events}?
	\item \textbf{RQ4:} What and why do they vary in \textbf{requirements engineering}? 
	\item \textbf{RQ5:} What and why do they vary in \textbf{quality assurance}?
\end{itemize}

\subsection{Case and Subjects Selection}

We selected the cases and subjects based on the availability and willingness of the interview partners. The cases, the specific projects where Scrum is applied, depend
on the study subjects, the interview participants, because they can only provide experience from their past or current projects. We also aimed to maximise variation by asking companies from different domains.

\subsection {Data Collection Procedure}
We conducted semi-structured interviews with the subjects about their most recent projects in which they applied Scrum. The guiding questions we used in the interviews are available 
in \cite{guidelines} and are aligned with common available Scrum checklists \cite{scrumchecklist}. Nonetheless, we did not use such checklists, because the reasoning 
behind the variations of Scrum is not in their scope.

We conducted all interviews by one of the authors as interviewer together with one company employee as interviewee. Within the interviews (1) we first explained the idea behind this work  to the participants. (2) We informed them that we handle their answers anonymously. (3) We gave them the interview questions and started discussing and answering. The result of the data collection were the final notes from each interview.

\subsection{Analysis Procedure}

We analysed the notes of the interviews purely qualitatively. First, we distilled categories with short answers into a table. For
example, we collected the Sprint length, the duration of the events or the team size for each case. Second, we extracted and
combined the answers for each of the research questions from the notes. We discussed and refined these answers among 
all researchers. For further discussion, we also checked possible connections between the asked questions and a mapping 
study for the usage of agile practices \cite{Diebold:2014:APP:2601248.2601254}. 

\subsection{Validity Procedure}

Our main action was to build and use the structured interview guideline to support the validity of the results. 
We selected the study subjects so as to avoid any interference between them. At the beginning, we stated the purpose of the interview, and we assured them that the results would be treated anonymously which gave the 
interviewees the freedom to give honest and open answers. As we did not record the interviews, we offered the interviewees the possibility to check the notes after the interview. 

All researchers read and discussed the interview notes as well as the extracted answers. For part of the table presenting the results, an independent re-extraction of the answers from the interview notes was conducted by two researchers to find and resolve discrepancies in the interpretation.

\section {Results}

\subsection{Case and Subject Description}

We conducted 10 interviews. The German companies of our interviewees cover a wide range from one very small start-up (4 employees) up to companies with around 130,000 employees. Six of the companies had a size between 100 and 350 employees. The remaining three companies are large corporations with several thousand employees. Most of the companies (except the smallest one) work and sell their products or services internationally. Besides the size and internationality of the organisations, we were interested in the different domains they were working in to further increase the variation. Yet, nine companies are working in different information systems domains and the other one in embedded systems. Our interviewees were all developers or development managers but are not necessarily representative for other Scrum teams in the same company.

\subsection{Overview}

We were able to give short answers for 14 aspects of the interview notes. These are shown in Table~\ref{tab:results}. We provide the team size (excluding Scrum Master and Product Owner), 
if tasks are outsourced, i.e. given to people outside of the team, and  if the team is at only one location. We show if there is a Scrum Master and a Product Owner. For all the event types,
we give the durations and, for the Daily Scrums, also if discussions are allowed beyond the answers to the three questions. For the Sprint planning, we report the Sprint lengths, if there
is a buffer in the plan, if there is a release plan and whether stories not completed in a Sprint are put back into the Product Backlog, split or continued. If we could not clearly determine the answer from the notes, we mark the cell with an ``?''.

\setlength{\tabcolsep}{6pt}
\begin{table}
\caption{Results of the Interviews}
\label{results}
\begin{threeparttable}
\begin{tabularx}{12.7cm}{L{.2cm}L{.8cm} L{.8cm} L{.8cm} L{1.8cm} L{2.0cm} L{1.4cm} L{1.4cm}}
\hline
   & Team   & Tasks & Team  & Scrum & Product  & \multicolumn{2}{c}{Daily Scrum}  \\ \cline{7-8}
  No. &  Size  & Out\-sourced&  Local & Master  & Owner & Duration & Discussions \\
\hline
 \textbf{1} & 3--7 			&  yes & yes 								& yes, is also project lead  		&  no  & 15 min, partly every second day & no    \\
 \textbf{2}  & 5 			&  yes & yes 								& yes, is also developer   			& yes, but also PO for the whole system  & 30 min, only when needed   & brief\\
 \textbf{3}  & 2--7  	& no 	& (yes)\tnote{1} 	& yes  &  yes & 15 min, but story related & no \\
 \textbf{4}  & 4--10 		& no 	&  ?							& yes, had additional tasks & no, divided between several people  & 15 min  & no \\
 \textbf{5}  & 20--25 split into 2 teams		& no 	& yes 								& no, divided between 3 people  & yes & 30 min & yes\\
 \textbf{6}  & 10 			& yes 	& yes 								& yes, is also developer & no  & 15 min  & yes \\
 \textbf{7}  & 2--4  		& no 	& yes 								&  yes  & yes, but is also developer & ? & ? \\
 \textbf{8}  & 5 + tester & yes & (yes)\tnote{2} & yes, is also team-leader & no & 15 min  & yes  \\
 \textbf{9}  & 10 & yes & yes & no & yes & 15 min & no\\
 \textbf{10} & 4 & no &  (yes)\tnote{3} & no & no, role split between architect and customer& 15 min & yes \\
 \hline
\end{tabularx}
\begin{tablenotes}
\item[1] Two adjacent rooms \item[2] same floor \item[3] If everybody is present
\end{tablenotes}
\end{threeparttable}
\bigskip

\begin{threeparttable}
\begin{tabular}{p{0.2cm}p{1.1cm} p{1.5cm} p{1.7cm} L{1.6cm} p{.7cm} p{1cm} p{1.5cm}}
\hline
&\multicolumn{3}{c}{Duration of Event} & Sprint& & Release& Incomplete \\ \cline{2-4}
No.&  Planning & Review & Retrospective & Length & Buffer &Plan  &Stories\tnote{4}\\
\hline
 \textbf{1} &  30 min  & \multicolumn{2}{c}{both together 1h} &  4 weeks &  10\%  & yes & back/split\\
 \textbf{2} &  1 day &  \multicolumn{2}{c}{both together 1 day} &  4 weeks & none & yes & back/cont.\\
 \textbf{3} & 1 day & 1h & 1.5h & 2 weeks & none & yes & back\\
 \textbf{4}  & 4--7h &  30 min &  1h &  2 weeks &  none & no& split\\
 \textbf{5} &   3h & 3h & ? & 3 weeks & 20\% & yes & cont.\\
 \textbf{6} &  ? & ? & ? & 4 weeks, $3 \times 2$ weeks & none &  yes & ?\\
 \textbf{7} &  1 day & 1 day & ?  &     1--4 weeks &   25\%  & no & split\\
 \textbf{8} & 1.5h &1.5h &1.5h& 2 weeks & no & no &split \\
 \textbf{9} & 4h & 1h--1.5h & 1h & 2 weeks & no & yes &split \\
 \textbf{10}&\multicolumn{3}{c}{all three together 3--4h}& 1--2 weeks & yes &  yes &  split\\
 \hline
 \end{tabular}
\begin{tablenotes}
\item[4] back = Back to the Backlog; cont. = Continue in the next Sprint; split = story is split up and unfinished work has to be planned again 
\end{tablenotes}
\end{threeparttable}
\label{tab:results}
\end{table}

\subsection{Team, Product Owner, and Scrum Master}

The team is a central part of Scrum and an important constraint is the size of the team. We found that several of the companies stretch the team size below and above the
recommended 3--9 people. Two of the companies have teams with only two members. Three companies work with teams of up to ten members; one of these even with more than 10 members.
The reason is that originally, there was a classical team of 25 people.

Some teams have dedicated experts for specific topics while others are generalists. The teams with experts explain their choice by the extraordinary technical depth and higher efficiency. The oddest case was a ``classical'' Scrum team and an additional team for writing specifications. The company considers this necessary, because they implement the core of a very large project with many other teams relying on them. The specifications team is responsible for acquiring information about all interfaces and from all the other teams. On the other hand, the teams with generalists argue that it reduces the problem of unavailable people and allows the team balance responsibilities better.

Most of the companies run cross-functional teams with all expertise necessary for the successful completion of the project. Two of the companies outsourced some aspects, e.g.\ UI design or manual testing.

 Ionel \cite{ft2008critical} found similar conclusions. He points out, as a possible cause for this, that smaller teams might work more effectively due to better communication, but the additional effort to coordinate a bunch of small teams increases significantly. So companies tend to increase the team size instead.


Half of the companies follow the standard idea of a \textbf{Product Owner} in their projects. Often, the Product Owner is a business analyst responsible for one or more teams (to reduce effort for communication between them). One company also had a hierarchy of Product Owners. Two companies even had a Product Owner directly from the customer. In contrast, in one company, the Product Owner was both, the business expert and the project manager. In one company, there were two Product Owners: one being the internal software architect and one being the external customer. Others reported that they either do not have a dedicated Product Owner at all and receive requirements directly from stakeholders or have a separate product management (department). Finally, in one company, a developer took this role because of the company size of four people.

It is interesting that not all interviewed companies had a Product Owner, as e.g. Moe and Dings{\o}yr~\cite{moe2008Scrum} stress that the Product Owner is crucial for the communication of the product vision.

Almost all interviewees stated that they use the role of the \textbf{Scrum Master} in some way. However, the implementation differs: Companies fill this role with an existing project manager or team lead, split it between project manager and software architect or have one of the developers as Scrum Master. Thus, the main difficulty seems to be that being Scrum Master for only one team is not a full-time job. Two companies report that having one of the developers as Scrum Master works well with a strong-minded and experienced developer, because such a person has a better insight into the technicalities of the project. This also increases his acceptance with the rest of the developers. In another company, where the Scrum Master is mainly a developer, the role degenerated to an event organiser. Only one company does not name a Scrum Master explicitly. In all companies without a dedicated Scrum Master, the costs seem to play a major role. They avoid reducing the overall capacity by assigning a developer as full-time Scrum Master.
Another possible cause for a ``shared" Scrum Master is presented by Moe and Dings{\o}yr~\cite{moe2008Scrum}. The role is shifted in the direction of a project manager, because the team members are working on many different projects simultaneously, and so the Scrum Master is also in charge of managing the progress of different projects.

\subsection{Sprint}

All interviewees reported that their companies run fixed-\textbf{length} Sprints. The length of these Sprints is mostly four weeks but some also used two or three weeks. The smallest company uses a fixed Sprint length in a project but varies over projects. They sometimes even run one-week Sprints. 
One company reported that they separate ``normal'' Sprints of four weeks from subsequent two-week Sprints for clean-up work. 
All interviewees reported that exceptions are rare, e.g. for public holidays. One company handles new product generations more flexibly but has fixed-length Sprints for established products. One interviewee reported that the Sprint is not shielded from outside changes to let the product management remove stories from or push stories into ongoing Sprints.

Most companies do not calculate a \textbf{buffer} in the work assigned to a Sprint. But two interviewees report that they only calculate with 80\% workload for the developers to account
for sick leave or uncertainties. One company uses a fixed 10\% buffer. Another reserves 25\% for bug fixing, grooming and any unforeseen work.
Another company has a varying buffer for technological risks.

\subsection{Events}

Although the \textbf{Daily Scrum} is a central means of communication in Scrum, we found that most of the companies do not follow \cite{sutherland2011Scrum}. Some companies
hold events of 30 minutes instead the 15 minutes. We also have results that the event is done every other day or only once a week, if there are not enough news.
One company allows members and Scrum Masters of other teams, who are responsible for interfaces, to be present at the Daily Scrum. This should make agreements on these interfaces easier. Furthermore, several interviewees reported that discussions are allowed during their Daily Scrums. The reason is that then they discuss issues relevant for everyone on the team and decisions can be made. One interviewee described that they hold the event structured according to the currently relevant User Stories and discuss them one by one because of higher efficiency.

A reason why the time span between Daily Scrums is increased might be the increased team size. As Ionel \cite{ft2008critical} stated, the increased event time also holds the risk of team members becoming uninterested. Companies might try to compensate for that by not holding the events daily, thus increasing the information content to keep it interesting for everyone.

All companies hold explicit \textbf{Sprint planning} events with varying topics from the current Sprint up to several Sprints. Most companies also follow the proposed structure of (1) fixing the stories for the Sprint and then (2) refining them into tasks. Six companies reported that they use planning poker for estimating User Stories. Sometimes, the planning poker sessions are held outside the planning. If they have very unclear stories, one company inserts a pre-planning phase of up to five days. Some companies skip the second part of the event. The small company even does not define any acceptance criteria because of the vaguely defined User Stories and the missing Product Owner taking care of that. Most companies reserve a whole day for the event and report that this investment pays off by accurate planning and estimates. One company reported only 30 minutes but we assume that they do not perform a proper planning.

All interviewees reported that their companies hold some kind of \textbf{Review} event. In several companies, other stakeholders are not always present
at the Review. In one company it is a means to get feedback about missing functionality from the Product Owner. Two companies have the strategy to conduct two
Reviews: One Review is internal with other developers reviewing the results. The second Review contains other stakeholders and in particular the customer. The reason is
that the team has the possibility to make smaller changes and corrections based on first feedback before the customer sees the increment.

Finally, the \textbf{Retrospective} is held in most companies. We found only one company that does not use Retrospectives at all. Another company holds them only rarely.
All interviewees report that the Retrospectives are held together or at least on the same day as the Review. In this event, however, only the team participates. The length of
the combined Review and Retrospective events range from 1 to 3.5 hours. Only the small company has a full day Review event.

\subsection{Requirements}

All interviewees use a \textbf{Product Backlog} as a central means of capturing requirements. As Scrum suggests, all companies keep the requirements in
the Product Backlog rather vague and high-level. One interviewee stated that one of the aims is to give an overview of the project. Several of the projects use \emph{JIRA}\footnote{\url{https://www.atlassian.com/software/jira}} for handling the Product Backlog. Microsoft Excel is in use alternatively.

The more concrete requirements in the \textbf{Sprint Backlog} are handled mostly as proposed in the Scrum Guide. Almost all interviewees described that the team
selects and refines requirements from the prioritised list in the Product Backlog. Only one company does not allow the team to decide on that but the Product Owner, architect
and project leader select and prioritise the requirements. This is a relic from the older, hierarchical development process.

The common way to specify requirements is by \textbf{User Stories}. In most companies, the team defines some more or less sharp acceptance criteria per User Story. Only one team in our study had  an actual Definition of Done in the Scrum sense.
In one company, the acceptance criteria are defined during Sprint planning. One company does not consequently use User Stories as means of describing requirements. They state that they use them only for a better understanding but not for all requirements. The reason is, again, the small size of the company and the missing Product Owner.
If a User Story cannot be completed in a Sprint, there are two strategies in the analysed companies: Either the whole User Story is pushed back to the Product Backlog and
reprioritised, or the team tries to split the User Story into something shippable now and tasks that are done in the next Sprint. The effort for a User Story is estimated either in
story points or person-hours. The companies using person-hours argue that they found story points too abstract and prefer to work with a more specific unit.

\subsection{Quality Assurance}

All interviewees described the usage of automated tests in their companies. They are usually part of a continuous integration and nightly builds. Only four companies explicitly
mention additional code reviews and automated static analysis. Four interviewees explicitly mentioned manual tests. One company emphasised that they also do a review of each 
User Story they define. Scrum has no explicit constraints on the used quality assurance techniques. 

The work of Fontana, Reinehr and Malucelli~\cite{fontana2014maturing} revealed that agile quality assurance can be added at any level of maturity. It is possible that the companies with less QA techniques in their development processes are at the beginning of their personal development in the agile world and so focus first on the essential parts, e.g.\ the involved customer or agile planning.

These quality assurance techniques are used to check the definition of done. This is an important concept in Scrum. The acceptance of a User Story with acceptance
criteria and a definition of done is practiced in the analysed companies. It varies, however, how strictly the Definition of Done is defined and who is deciding acceptance.
Several interviewees stated that the acceptance criteria are not clearly specified. In some companies, the Product Owner decides if the User Story is accepted. If there is no
Product Owner, the team makes this decision.

\subsection{Evaluation of Validity}

The interview guidelines proved to be helpful for focussing during the interviews but even more so during analysis as the interviews were not audio taped. We used the categories as a guiding structure in the analysis and write-up. Additionally, the interview guidelines reduced the risk of a misinterpretation and increased the objectivity of the notes taken, because it was always possible to fall back to the basic question in the guideline. The remaining threat of subjective filtering by the interviewers is in our opinion negligible.
 We did not notice any major misunderstandings. For example, some interviewees were not directly aware of what the three questions
in the Daily Scrum are but a short explanation could resolve this. Furthermore, we had the impression that the assurance of anonymity
led the interviewees to answer freely and openly.

The independent re-extraction of several of the answers in the main results table (Table~\ref{tab:results}) revealed few differences in our interpretation. For example,
we judged differently under which circumstances we describe a case as having a Scrum Master or Product Owner. A discussion resolved these differences. For the roles,
we decided to stick to the Scrum Guide and not accept a Scrum Master or Product Owner, if the role is shared by several people. Therefore, we are confident that the contents
of the table are valid. For the further textual results descriptions, we cannot rule out that there are smaller misinterpretations. Yet, all researchers reviewed these parts and we
discussed unclear issues.

Despite we only interviewed German development teams, we believe that our qualitative results should be well generalisable for other companies applying Scrum, especially in information systems. We expect the variations and reasons will occur in other companies, maybe among others. Still, there might be a a cultural impact, which we are investigating in this study.

\section {Related Work}


In contrast to our purely outside view on the topic, Kniberg \cite{kniberg2007Scrum} reports from his experience how Scrum and XP is used in the real world. He discusses essential parts of the process in detail, including some of the alterations we have seen in the interviews.

Kurapati, Manyam and Petersen \cite{kurapati2012agile} did an extensive survey of agile practices. Among other topics, they also looked into compliance to the Scrum framework. But while Kurapati et al. stopped at the level of how many Scrum practices were used, we go one step further and investigate in detail which practices are used and how and why they are altered.

Moe and Dings{\o}yr~\cite{moe2008Scrum} examined the team effectiveness effects of Scrum. They formulated the alterations of the company involved in the case study as problems. This is a different perspective compared to our work, but it still shows which kind of alterations are made and why.


Dorairaj, Noble and Malik~\cite{dorairaj2012understanding} studied the behaviour of distributed agile development teams. They focused on the dynamics of cooperation in the teams and presented six strategies the teams adopt to make up for the difficulties in communication in distributed teams.
Their data also provided support for our results concerning the topic of team size as a frequently violated Scrum rule and the almost complete commitment to the Sprint length of 2--4 weeks.


Barabino et al.~\cite{barabino2014agile} conducted a survey on the use of agile methodologies in web development. From the Scrum practices, they found that the Daily Scrum is used most often and that aspects connected to the releases, like continuous delivery, are taken care of less. This matches our results, as we see that the daily events are used by all of our interviewees with little changes and process parts, e.g.\ the release plan, are used seldom and only in a very vague form.

%
%


Ionel \cite{ft2008critical} discusses key features of Scrum, like the team size or the Sprints, and potential effects of deviations from these key features. For example, he states that a team of more than 10 people will have increasing difficulties in communicating and implementing changes. Yet, splitting a larger team into several smaller teams leads to a large coordination effort (Scrum-of-Scrums).

Fontana, Reinehr and Malucelli~\cite{fontana2014maturing} thought about what defines maturity in agile development. They argue that maturity in agile is not about following a predefined path but to find what fits your agile development style.  They still see some essential agile practices enforced by most of the mature agile users. While this is on a higher abstraction level then our work, the effect of maturity stated here might be the reason of some of the changes we see in Scrum.

\section{Conclusions and Future Work}
Based on the ten interviews performed with different companies about their applications of Scrum, we can confirm the statement of Schwaber 
that most often it is not used as proposed. Our results show that (1) none of the companies conforms to the Scrum Guide (only one is close) and 
(2) there is at least one company deviating from the standard for each aspect. Additionally, the results of the interviews gave us several reasons 
for these variations. In some cases, we found pragmatic justifications such as ``the team found it more efficient". For example, short discussions 
during the Daily Scrum seem to be useful in some companies. Other deviations seem more like a legacy from more hierarchical, non-agile processes.
For example, one company has a specification team and an implementation team as well as the Scrum Master role split between a project leader
and a chief architect. 

In addition to the comparison with related work, we conclude by relating our results to the overall results of a mapping study on agile practices~\cite{Diebold:2014:APP:2601248.2601254}.
The results concerning the \textbf{Sprints} and their lengths showed similar results as in the literature: all companies are using a time box. The partial variation of the Sprint length is also similar in literature. 
Of the events performed \textbf{Sprint Planning} is most often mentioned in literature, followed by the \textbf{Retrospective} and less often, the \textbf{Review}. In contrast, our results show the opposite: Retrospectives are used less by the interview partners. 
The common use of  \textbf{Daily Scrums} is confirmed by our results and the mapping study and the few deviations in the event durations can be found in literature too.  

Regarding requirements, again a similarity to~\cite{Diebold:2014:APP:2601248.2601254} can be seen, as \textbf{User Stories} are used by all of the interview partners. They only vary the way of writing. 
Additionally, our results show the different variations of dealing with the concepts of Product Backlog and Sprint Backlog.
The mapping study covers all agile methods, also XP including the on-site customer which is rarely used. This explains the deviation from our \textbf{Product Owner} results. Product Owners are often used but frequently in slightly adapted ways. 
The QA aspects show the largest deviation between the mapping study and our interview results, because literature often mentions explicit the absence  of pair programming, whereas our results give more details about which QA practices are used within Scrum. This matches the partial usage of QA practices reported in~\cite{Diebold:2014:APP:2601248.2601254}.

Based on our results we would like to extend this case study to companies with more varying background. Additionally, it would be helpful to interview companies from different countries. Then a detailed comparison with domain data of~\cite{Diebold:2014:APP:2601248.2601254} would be possible, and we might be able to give practitioners guidance on when to vary which aspects of Scrum. 


\bibliographystyle{plain}
\bibliography{xp15}

\end{document}